\documentclass[trackchanges]{aastex701}
\usepackage{multirow}

\begin{document}

\title{Performance Simulations for KOLA: Achieving High-Resolution, Visible-Light AO Correction over a 1' Field}

\author[orcid=0000-0001-9992-7022]{Brianna Peck}
\affiliation{Department of Astronomy, University of California, Berkeley, CA, USA}
\email{bpeck114@berkeley.edu}  

\author[orcid=0000-0001-9611-0009]{Jessica R. Lu} 
\affiliation{Department of Astronomy, University of California, Berkeley, CA, USA}
\email{jlu.astro@berkeley.edu}

\author[orcid=0000-0002-3523-577X]{Lianqi Wang} 
\affiliation{TMT International Observatory, Pasadena, CA, USA}
\email{lianqiw@tmt.org}

\author[orcid=0000-0003-4700-428X]{Brooke DiGia} 
\affiliation{Department of Astronomy, University of California, Berkeley, CA, USA}
\email{brooke_digia@berkeley.edu}

\author[orcid=0000-0002-5884-7867]{Richard Dekany}
\affiliation{Caltech Optical Observatories, California Institute of Technology, Pasadena, CA, USA}
\email{rgd@astro.caltech.edu}

\author[orcid=0000-0001-9954-4398]{Antonin H. Bouchez}
\affiliation{W. M. Keck Observatory, Waimea, HI, USA}
\email{abouchez@keck.hawaii.edu}

\author[orcid=0000-0002-1646-442X]{Peter Wizinowich}
\affiliation{W. M. Keck Observatory, Waimea, HI, USA}
\email{peterw@keck.hawaii.edu}

\author[orcid=0000-0001-6205-9233]{Maxwell A. Millar-Blanchaer}
\affiliation{Department of Physics, University of California, Santa Barbara, CA, USA}
\email{maxmb@ucsb.edu}

\author[orcid=0000-0002-8462-0703]{Mark Chun}
\affiliation{Institute for Astronomy, University of Hawai‘i at Mānoa, Hilo, HI, USA}
\email{markchun@hawaii.edu}

\author[0000-0002-1954-4564]{Philip Hinz}
\affiliation{Laboratory for Adaptive Optics, University of California, Santa Cruz, CA, USA}
\email{phinz@ucsc.edu}

\author{Charles-Antoine Claveau}
\affiliation{Department of Astronomy, University of California, Berkeley, CA, USA}
\email{caclav@berkeley.edu}

\begin{abstract}

We present performance simulations for a proposed visible-light, multi-conjugate adaptive optics system for the 10-meter W. M. Keck I telescope that aims to deliver near diffraction-limited angular resolution at optical wavelengths. 
Our proposed architecture, the Keck Optical Laser Guide Star Adaptive Optics System (KOLA), combines multiple laser guide stars (LGS) and deformable mirrors to enable wide-field correction across a 60" field of view. 
Simulations were conducted using the open-source Multi-Threaded Adaptive Optics Simulator (MAOS), which we validated against on-sky data for the current Keck I adaptive optics system. 
We evaluated KOLA performance across a range of design parameters and report key point spread function metrics, including Strehl ratio, full width at half maximum, and encircled energy radius. 
Example science-driven requirements include resolving black hole spheres of influence, probing crowded stellar fields, and imaging protoplanetary disks. 
Trade studies on actuator count and laser guide star configuration help inform future design decisions. We present a nominal KOLA design (10 LGS, 3 tip-tilt natural guide stars (TTNGS), and 3600 actuators on the adaptive secondary mirror). 
Performance simulations show a $15\,\text{mas}$ angular resolution with a Strehl ratio of 34\% at $652\,\text{nm}$ on-axis. 
More work is needed to explore alternative LGS/TTNGS asterisms, optimize conjugation heights for high-altitude deformable mirrors, and test performance under poorer seeing conditions.

\end{abstract}



\section{Introduction}

Atmospheric turbulence degrades the angular resolution of large ground-based optical and infrared observatories, necessitating the use of adaptive optics (AO) to correct wavefront distortions and restore diffraction-limited performance. 
Because tip-tilt natural guide stars (TTNGS) are limited in brightness and sky distribution, even AO systems equipped with artificial laser guide stars (LGS) must balance sky coverage, image quality, and corrected field of view.

The twin 10-meter telescopes at the W. M. Keck Observatory each operate a single-conjugate adaptive optics (SCAO) system using a single deformable mirror (DM) and either a natural or sodium laser guide star, feeding high-resolution near-infrared instruments such as NIRC and NIRC2 \citep{2006PASP..118..297W}. 
Under good seeing conditions, the Keck LGS SCAO system achieves near-diffraction-limited performance in the near-infrared (e.g., K-band Strehl ratios of 30-40\%) \citep{2006PASP..118..310V}. However, performance declines rapidly at shorter wavelengths and with increasing angular separation from the LGS due to limitations in tip-tilt sensing, finite laser return, and anisoplanatic errors \citep{2006SPIE.6272E..31V}. 
Additionally, visible-light corrections are particularly sensitive to atmospheric turbulence, the focal anisoplanatism introduced by a single sodium LGS (i.e., the ``cone effect"), and the intrinsically narrow corrected field of view of SCAO systems. 

The Keck All-sky Precision Adaptive Optics (KAPA) upgrade on the Keck I telescope represents a major step toward tomographic AO for the Keck Observatory. 
KAPA will deploy a narrow LGS asterism produced by splitting a single sodium LGS into four beacons, sensing them simultaneously with a new real-time controller capable of multi-laser tomographic reconstruction \citep{2020SPIE11448E..0EW}. 
KAPA will also implement multi-star TTNGS sensing, combining several faint stars to improve low-order correction and expand sky coverage. 
Following KAPA, the planned Ground-Layer Adaptive Optics (GLAO) upgrade for Keck aims to correct only the dominant low-altitude turbulence over a large field of view \citep{2018SPIE10703E..0NL}.
Together, KAPA and GLAO serve as stepping-stones toward a full high-resolution visible-light system for Keck.

\begin{figure*}[t!]
    \centering
    \includegraphics[width=\linewidth]{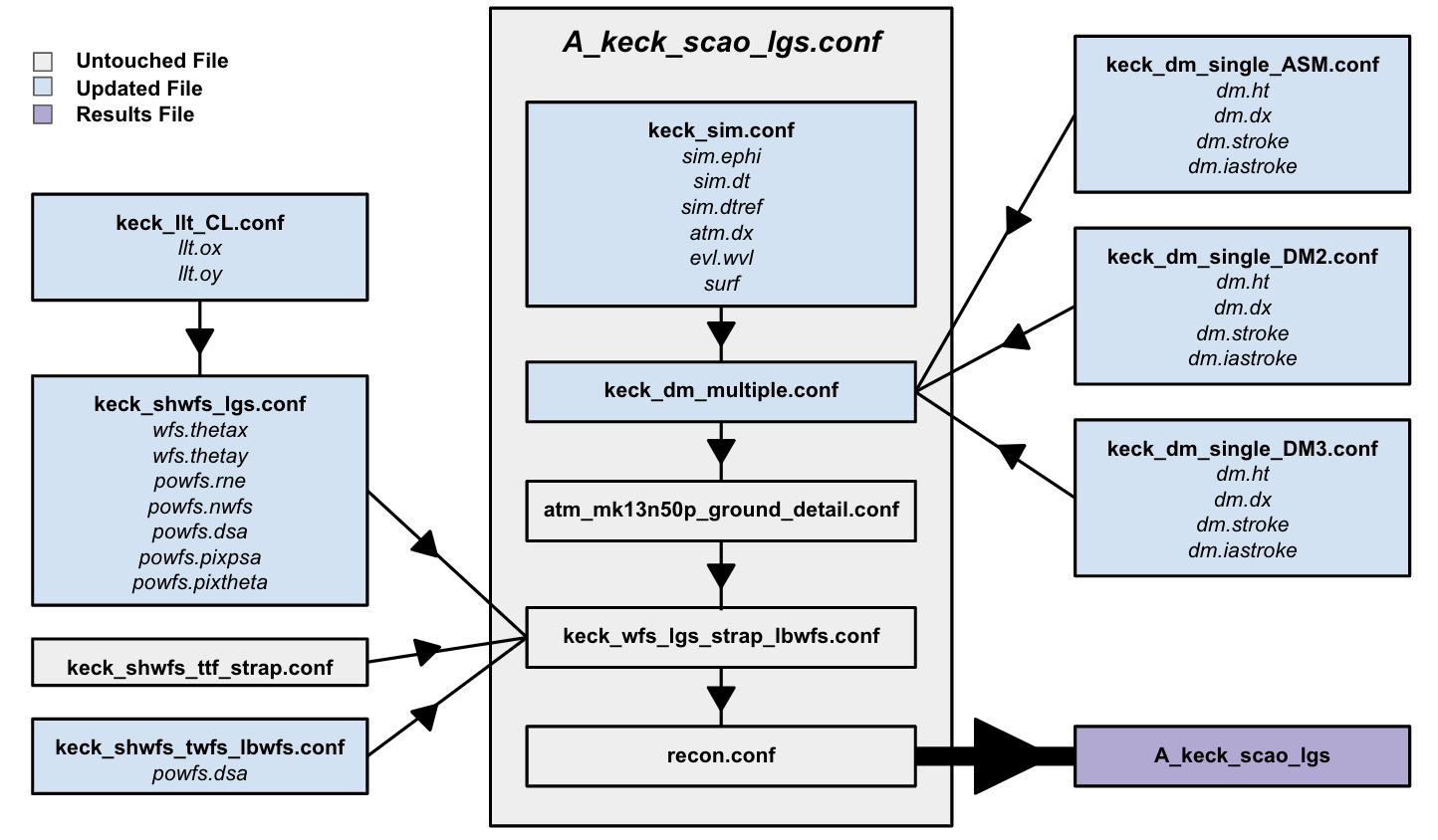}
    \caption{Structure of the MAOS simulation package. A master configuration file defines global simulation settings and sequentially calls the configuration files for the DMs, atmosphere, and wavefront sensors. This hierarchical structure enables coordinated control of all simulation components while allowing global parameter changes without modifying each configuration file individually. Simulation outputs (e.g., PSFs, telemetry, and wavefront error data) are written to a designated results directory.}
    \label{fig:maos_chart}
\end{figure*}

The 2035 Keck Strategic Plan identifies visible-light AO as a critical goal, emphasizing the need for a corrected field of 30''-60'' in diameter, higher sky coverage, and diffraction-limited resolution to complement Extremely Large Telescopes (ELTs) and space telescopes \citep{Keck2035}.
Similar capabilities are under development at the 8-meter Very Large Telescope (VLT) through the MAVIS (Multi-conjugate Adaptive-optics Visible Imager-Spectrograph) project, which passed its Preliminary Design Review in 2023 and is now in its Final Design Review (Phase B) phase \citep{2022SPIE12184E..3SC}.
Motivated by these developments, this paper explores possible designs for a comparable visible-light system for the Keck I telescope. 

Achieving diffraction-limited performance at visible wavelengths over a moderate field of view on Keck requires four major upgrades.
First, multiple independent LGS for laser-tomographic AO (LTAO) are needed to mitigate focal anisoplanatism by sampling a larger fraction of the atmospheric volume above the pupil \citep{1990A&A...235..549T}. 
While KAPA implements a four-spot LGS constellation, a next-generation system will require multiple full-power laser launch telescopes to achieve the necessary return flux and tomographic sampling. 
Second, multi-conjugate adaptive optics (MCAO) using several DMs conjugated to different altitudes enables a wider and more uniform corrected field of view \citep{1988ESOC...30..693B}.
Third, an adaptive secondary mirror (ASM) increases throughput and reduces thermal background \cite{2020SPIE11448E..5UH}. 
Fourth, multi-wavefront-sensor tip-tilt measurements, combining several faint TTNGS, can provide robust low-order sensing and improve sky coverage \citep{Reeves_2016}. 
Together with faster real-time control, these upgrades offer a path to overcoming the limitations of visible-light SCAO. 

In this work, we present the Keck Optical Laser Guide Star Adaptive Optics System (KOLA), a visible-light MCAO concept that integrates laser tomography, an adaptive secondary mirror, and additional high-altitude DMs. 
KOLA is designed to deliver wide-field, near-diffraction-limited imaging at visible wavelengths, enabling science cases inaccessible to current ground- and space-based facilities.
To guide the system design, we perform trade studies assessing KOLA's sensitivity to actuator density on the ASM and the number of independent LGS.
These studies identify the most critical design parameters and outline a roadmap toward a future visible-light MCAO facility for Keck. \\ 

This paper is organized as follows: Section \ref{sec:MAOS: Multi-Threaded Adaptive Optics Simulator} describes the MAOS simulation package and our validation of Keck SCAO simulations against on-sky performance. 
Section \ref{sec:Science Requirements and Cases for KOLA} summarizes science drivers for visible-light MCAO.
Section \ref{sec:KOLA Performance Simulations} presents the KOLA configuration, trade studies, and performance results.
Section \ref{sec:Discussion and Conclusions} discusses design implications and future work. \\

\section{MAOS: Multi-Threaded Adaptive Optics Simulator}
\label{sec:MAOS: Multi-Threaded Adaptive Optics Simulator}

\begin{deluxetable}{|c|l|}[t!]
\tablecaption{Top-level performance requirements for KOLA.}
\label{tab:science_requirements}
\tablehead{
\colhead{\textbf{Number}} &
\colhead{\textbf{Requirement}}
}
\startdata
1 &
\begin{tabular}{l}
The AO system will deliver AO-corrected light to \\
science instruments from 350--2500 nm.
\end{tabular}
\\
\tableline
2 &
\begin{tabular}{l}
The AO system will deliver a spatial resolution that \\
is within 10\% of diffraction limited down to 18 mas.
\end{tabular}
\\
\tableline
3 &
\begin{tabular}{l}
The AO system will have a sky coverage $>$50\% at the \\
Galactic poles with a maximum 50\% encircled energy \\
radius of 50 mas.
\end{tabular}
\\
\tableline
4 &
\begin{tabular}{l}
The corrected field of view will equal or exceed 60'' \\
in diameter.
\end{tabular}
\\
\tableline
5 &
\begin{tabular}{l}
For observations at 650 nm using TTNGSs with H$<$12 mag \\
within a 60'' radius, the system will deliver a Strehl \\
ratio of at least 30\% over a 20'' diameter field of view \\
and at least 15\% over a 30'' diameter field of view.
\end{tabular}
\\
\tableline
6 &
\begin{tabular}{l}
AO correction will be tunable to deliver a Strehl ratio \\
$\geq$ 0.5 over 3'' radius fields of view at 500 nm \\
for on-axis objects with V$<$6 mag serving as natural \\
guide stars, or for faint objects with 3 TTNGS stars of \\
H$<$12 mag within 15'' radius.
\end{tabular}
\\
\enddata
\end{deluxetable}

End-to-end physical optics simulations are essential for accurately capturing the performance of complex SCAO and MCAO systems. 
We modeled our proposed KOLA system using the open-source Multi-Threaded Adaptive Optics Simulator (MAOS)\footnote{MAOS is available on GitHub at \url{https://github.com/lianqiw/maos.git}.}, originally developed for Thirty-Meter Telescope AO studies \citep{wang2018maos}. 
MAOS propagates wavefronts through evolving frozen-flow atmospheric phase screens, applies the specified AO correction, and produces the resulting focal plane point-spread functions (PSFs) along with the residual wavefront error.
The package is highly configurable, with nearly 500 adjustable parameters for end-to-end AO simulations. 
Figure \ref{fig:maos_chart} shows how MAOS organizes its configuration files and highlights the parameters we modified for our KOLA performance trade studies. 

With the implementation of GPU acceleration, we were able to run these studies on the NERSC (National Energy Research Scientific Computing Center) supercomputing facility. 
This capability allowed us to explore a much broader parameter space than would be feasible with CPU-only simulations. 

To validate the MAOS simulation package results, we first performed a simulation of the current Keck SCAO system. 
The modeled Keck AO system includes a single-conjugate (single-DM) AO system equipped with a sodium LGS for high-order correction and a fainter natural guide star used to measure tip-tilt and focus. 
This MAOS simulation of the SCAO system is fully described in Section \ref{appenidx:Validation of Current Keck AO System in MAOS}. 
The validation results show that to match on-sky Strehl ratio and FWHM measurements, two unknown error terms must be added: 
(1) a high-order wavefront error attributed to non-common path optical aberrations and 
(2) a tip-tilt and jitter power spectrum attributed to wind shake and AO system vibrations. 
More details on the amplitude of the two error terms can be found in Section \ref{appendix:MAOS Parameters for KOLA Performance Simulations}.
The exact origins and mitigation of these two error terms is still under investigation. 
For the purposes of KOLA MAOS simulations, we assume that these terms will be understood and corrected prior to KOLA installation.

\section{Science Requirements and Cases for KOLA}
\label{sec:Science Requirements and Cases for KOLA}

The performance requirements for KOLA are motivated by both general AO benchmarks and the specific needs of science cases that demand near-diffraction-limited visible-light imaging. 
Table \ref{tab:science_requirements} summarizes the top-level science requirements: continuous AO correction across the visible to near-infrared band, image quality that tracks the diffraction limit down to blue wavelengths, a uniformly corrected field large enough for mapping and astrometry, and sky-coverage targets that make these capabilities usable at high Galactic latitude. 

\begin{deluxetable}{c l l}
\tablecaption{A list of preliminary example science cases for KOLA which require higher spatial resolutions and larger fields of view than can be achieved with current facilities (Keck AO, HST, JWST).}
\label{tab:science_cases}
\tablehead{
\colhead{\textbf{Number}} &
\colhead{\textbf{Science Case}} &
\colhead{\textbf{Measurement Requirements}}
}
\startdata
1 &
\begin{tabular}{l}
Image and measure spectra of individual stars \\
surrounding the supermassive black hole in the \\
Milky Way, M31, and dwarf galaxies in the Local Group.
\end{tabular}
&
\begin{tabular}{l}
Probe stellar populations within the sphere of \\
influence($1.6$ AU in M31). Requires spatial  \\
resolution of $\leq 20~\mathrm{mas}$.
\end{tabular}
\\
\tableline
2 &
\begin{tabular}{l}
Image and characterize accreting protoplanetary disks.
\end{tabular}
&
\begin{tabular}{l}
Resolve $10~\mathrm{AU}$ structures in Orion ($400$ pc), \\
requiring spatial resolution of $\leq 20~\mathrm{mas}$.
\end{tabular}
\\
\tableline
3 &
\begin{tabular}{l}
Find and weigh isolated black holes and neutron stars \\
in the Galactic Bulge through gravitational lensing.
\end{tabular}
&
\begin{tabular}{l}
\textbf{Case A:} Astrometric precision $<60$ mas \\
on $R=22$ stars at $10$ kpc; field of view $\geq 30''$.\\[2pt]
\textbf{Case B:} Resolve lensed images separated by  \\
$9$ mas; requires resolution of $\leq 18$ mas.
\end{tabular}
\\
\tableline
4 &
\begin{tabular}{l}
Monitor and study transient storms and volcanic activity \\
on Solar System planets, moons, and small bodies.
\end{tabular}
&
\begin{tabular}{l}
Resolve $500$ km features on Uranus/Neptune  \\
and$25$ km features on Mars/Venus  \\
($<23$ mas). Field of view $>50''$ to image  \\
Jupiter, Mars, Venus, and comets. Wavelength \\
coverage: $380$–$900$ nm.
\end{tabular}
\\
\tableline
5 &
\begin{tabular}{l}
Monitor planetary activity, accretion, weather, and \\
habitability tracers (UV, H$\alpha$, atomic/molecular species).
\end{tabular}
&
\begin{tabular}{l}
Achieve contrasts of 5–9 mag at 656 nm, \\
10–14 mag at 350 nm over $\Delta\lambda=350$–$2300$ nm.
\end{tabular}
\\
\tableline
6 &
\begin{tabular}{l}
Find and weigh black holes and neutron stars in binary \\
systems within $3$ kpc.
\end{tabular}
&
\begin{tabular}{l}
Astrometric wobble detection at $<100$ $\mu$as; \\
RV precision $<5$ km s$^{-1}$ at $R=21$ mag.
\end{tabular}
\\
\enddata
\end{deluxetable}

Table \ref{tab:science_cases} ties the science requirements from Table \ref{tab:science_requirements} into possible science observations. 
Crowded-field stellar imaging near supermassive black holes and in the Local Group is one of the drivers toward visible-light angular resolution with high Strehl ratios. 
Protoplanetary disk substructures set the need to resolve $\sim10\,\text{AU}$ features in nearby star-forming regions (like the Orion and Taurus cluster). 
Astrometric searches for isolated stellar-mass black holes motivate both a wide, uniformly corrected field (to secure a dense reference grid) and tight, field-stable PSFs. 
Solar system weather and small-body science push for large fields and broad visible coverage to capture key molecular/atomic bands in a single observation. 

These requirements ensure that KOLA does more than match Keck's current near-diffraction-limited infrared SCAO system, it extends Keck's AO reach into the visible, enabling a combination of field size, spatial resolution, and sky coverage that neither future ground-based Extremely Large Telescope systems nor space telescopes (HST or JWST) can currently achieve. 
The science requirements for KOLA parallel the approach taken by the VLT/MAVIS visible-MCAO project, which targets $<20\,\text{mas}$ resolution at $550\,\text{nm}$ across a $30''$ field with at least 50\% sky coverage at the Galactic Pole \citep{2020arXiv200909242M}.

Simulations of the MAVIS system show that their visible-light MCAO can achieve $\sim50\,\mu\text{as}$ astrometric precision for stars brighter than 19th magnitude. With that performance, they were able to detect the velocity-dispersion signature of a $1,500\,M_{\odot}$ black hole in the dense core of a globular cluster. 
This case connects to KOLA's goals: achieving stable, diffraction-limited PSFs across a $\geq 60''$ field and maintaining the needed Strehl ratios at visible wavelengths needed for precision astrometry in crowded fields \citep{2021MNRAS.507.2192M}. 
With Keck's larger aperture and extended field, KOLA would hopefully build on this capability to reach fainter stars and denser environments.

\subsection{Examples of Science Cases for KOLA}
\label{subsec:Examples of Science Cases for KOLA}

\begin{figure}[t!]
    \centering
    \includegraphics[width=\linewidth]{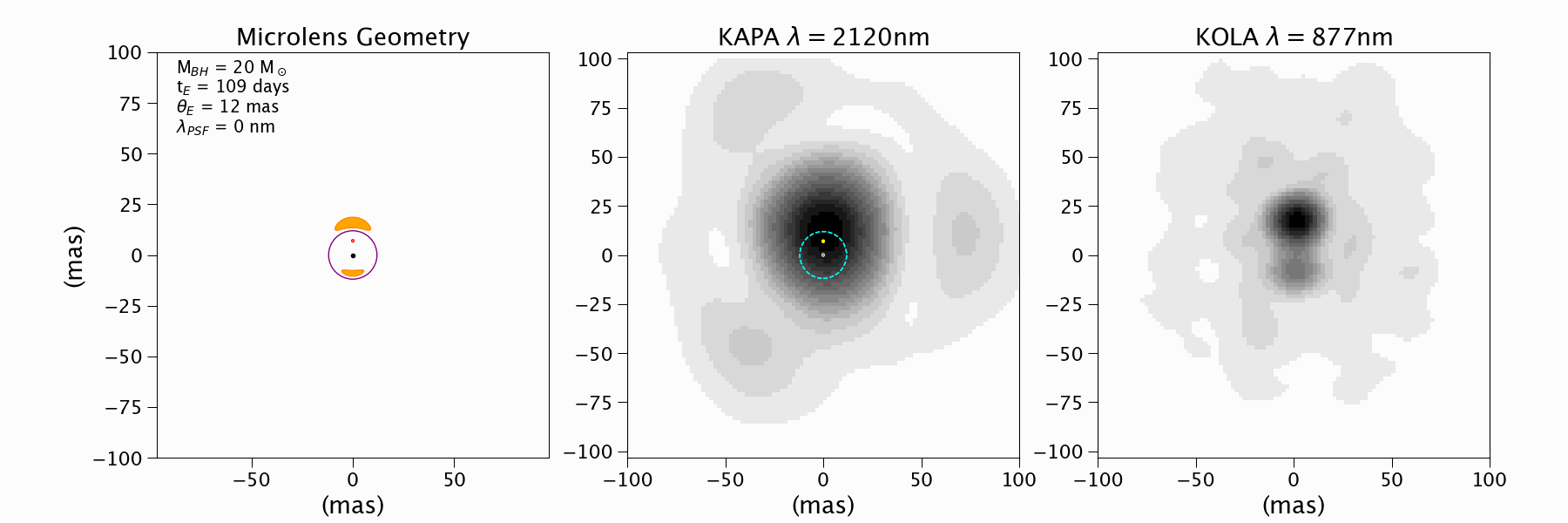}
    \caption{Simulated astrometric microlensing event comparing performance with the current Keck AO system (middle) and with KOLA (right). KOLA's higher angular resolution enables the two lensed stellar images to be separately resolved during the event, allowing direct measurement of the lensing geometry. This capability will enable systematic detection and weighing of stellar-mass black holes in the Milky Way, one of KOLA's key science drivers.}
    \label{fig:microlens}
\end{figure}

\begin{figure}[t!]
    \centering
    \includegraphics[width=0.5\linewidth]{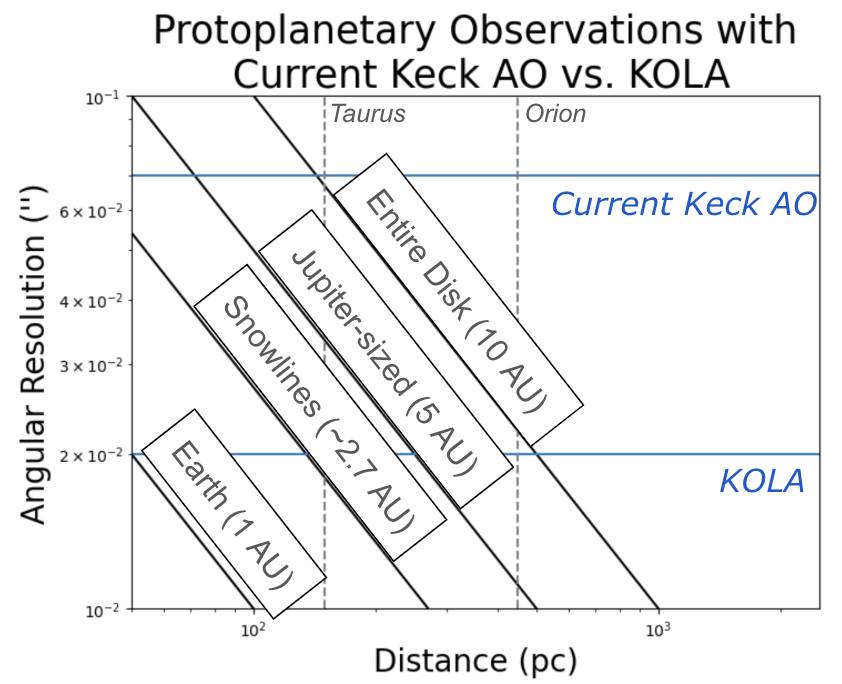}
    \caption{Angular resolution required to resolve key features in protoplanetary disks as a function of distance. At $150\,\text{pc}$ (Taurus cluster), current Keck AO can detect large-scale disk strucutre, while KOLA will reach $\sim 3\,\text{AU}$ resolution, sufficient to probe the terrestrial planet-forming region. At $400-450\,\text{pc}$ (Orion cluster), KOLA will still achieve $\sim10\,\text{AU}$ resolution, allowing direct imaging of gaps, structure, and snowlines sculpted by giant planets. This level of detail is inaccessible to current ground-based AO or space telescopes.}
    \label{fig:proto-disk}
\end{figure}

A compelling science case for KOLA is its ability to identify and weigh isolated stellar-mass black holes within $\sim1\,\text{kpc}$ through astrometric microlensing \citep{2016ApJ...830...41L,2022ApJ...933L..23L}.
By resolving the two lensed stellar images, KOLA can instantly constrain the lens mass.
Figure \ref{fig:microlens} compares simulated microlensing events observed with the soon-to-be KAPA system ($2120\,\text{nm}$) and with KOLA ($877\,\text{nm}$).

Planet formation offers another example of where KOLA's visible-light correction will be transformative.
At distances of a few hundred parsecs, KOLA can resolve substructures within protoplanetary disks on $3-10\,\text{AU}$ scales, precisely the regime where rings, gaps and snowlines trace ongoing planet formation \citep{2018ApJ...869L..41A}. 
Figure \ref{fig:proto-disk} illustrates how the resolution required to detect such features scales with distance.
At $150\,\text{pc}$ (Taurus cluster), KOLA will access the terrestrial planet-forming region; at $450\,\text{pc}$ (Orion cluster), KOLA will probe the regime of giant planet formation inaccessible to current AO or space facilities. 

\begin{figure}[t!]
    \centering
    \includegraphics[width=\linewidth]{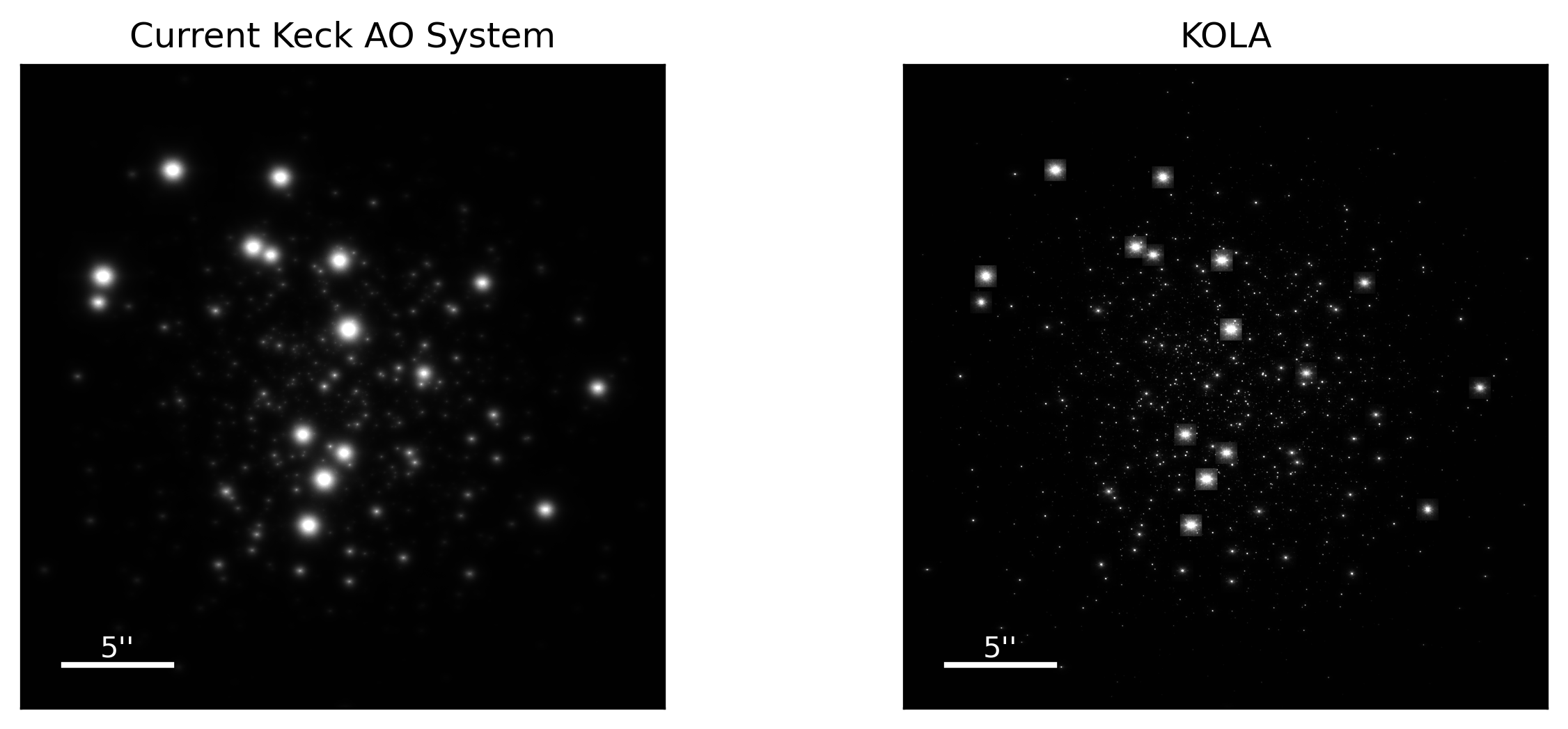}
    \caption{Simulated imaging ($810\,\text{nm}$) of a young stellar cluster at $4\,\text{kpc}$ using MAOS PSFs. Left: performance with the current Keck AO system. Right: performance with KOLA. The KOLA simulation resolves significantly more faint stars and disentangles crowded sources that remain blended in current AO imaging. The 30'' field of view is shown on a logarithmic scale, with stellar magnitudes assigned using the SPISEA package. This demonstrates KOLA's capability to identify faint populations in dense environments across the Galaxy. Due to the finite size of the PSFs produced by MAOS, clipping occurs for the KOLA image. However, it is still possible to see the higher angular resolution capabilities.}
    \label{fig:cluster}
\end{figure}

Also, KOLA's performance extends to dense stellar environments where precision photometry and astrometry are limited by crowding. 
Figure \ref{fig:cluster} shows a simulated $30''$ field of view of a $4\,\text{kpc}$ cluster whose magnitudes were generated with SPISEA \citep{2020AJ....160..143H} using a broken-power-law initial mass function ($\alpha = [-1.3, -2.3, -2.3]$), a total mass of $10^{5}\,M_{\odot}$, and an age of $5\times10^6\,\text{years}$. 
Positions were distributed using a 2D gaussian radial profile.
With current AO, only the brightest stars are recovered; with KOLA's simulated PSFs across a field of view, the cluster improves in resolution, revealing faint sources and enabling higher-precision photometry and astrometry. 
Together, these figures begin to highlight the breadth of science unlocked by KOLA, from black hole astrometry to planet formation to stellar population studies, all unified by diffraction-limited wide-field performance across the visible and near-infrared spectrum.

\section{KOLA Performance Simulations}
\label{sec:KOLA Performance Simulations}

As described in Section~\ref{sec:MAOS: Multi-Threaded Adaptive Optics Simulator} (and detailed in Section \ref{appendix:MAOS Parameters for KOLA Performance Simulations}), we first built and validated a baseline MAOS model of the current Keck LGS SCAO system.
After confirming that this configuration reproduced on-sky performance, we modified it to represent a visible-light, laser-tomographic MCAO system incorporating an ASM.
This KOLA configuration was then used to explore end-to-end performance as a function actuator count on the ASM and the number of sodium LGS (Section \ref{subsec:Actuator and LGS Trade Study Configuration}).

Each simulation runs in closed-loop at $1.5\,\text{kHz}$ for $6,400$ time steps, corresponding to approximately $4.3\,\text{seconds}$ of on-sky correction.
PSFs are computed at nine wavelengths from $432\,\text{nm}-2200\,\text{nm}$, both on-axis and at field locations out to $60''$ off-axis.
Performance is quantified using Strehl ratio, the empirical FWHM, and encircled energy (EE).
The Strehl ratio is computed as the ratio of the peak value of each simulated PSF to the peak of a diffraction-limited PSF at the same wavelength.
For the empirical FWHM, PSF pixels are sorted by radial distance from the center, the radius at which the intensity falls to $50\%$ of the peak is identified, and the FWHM is defined as twice this radius. 
The residual wavefront error (WFE) is also decomposed into tip-tilt and higher-order terms to identify the dominant contributors to performance degradation.
Further details on the KOLA configuration is presented in results Section \ref{sec:Setup and MAOS Configuration}, and the simulations results are summarized in Section \ref{subsec:Performance Simulation Results}.
Together, these simulations establish the performance benchmarks that inform KOLA's design requirements and define the nominal configuration described later in Section \ref{sec:Nominal Design for KOLA}.

\begin{figure}[t!]
    \centering
    \includegraphics[width=\linewidth]{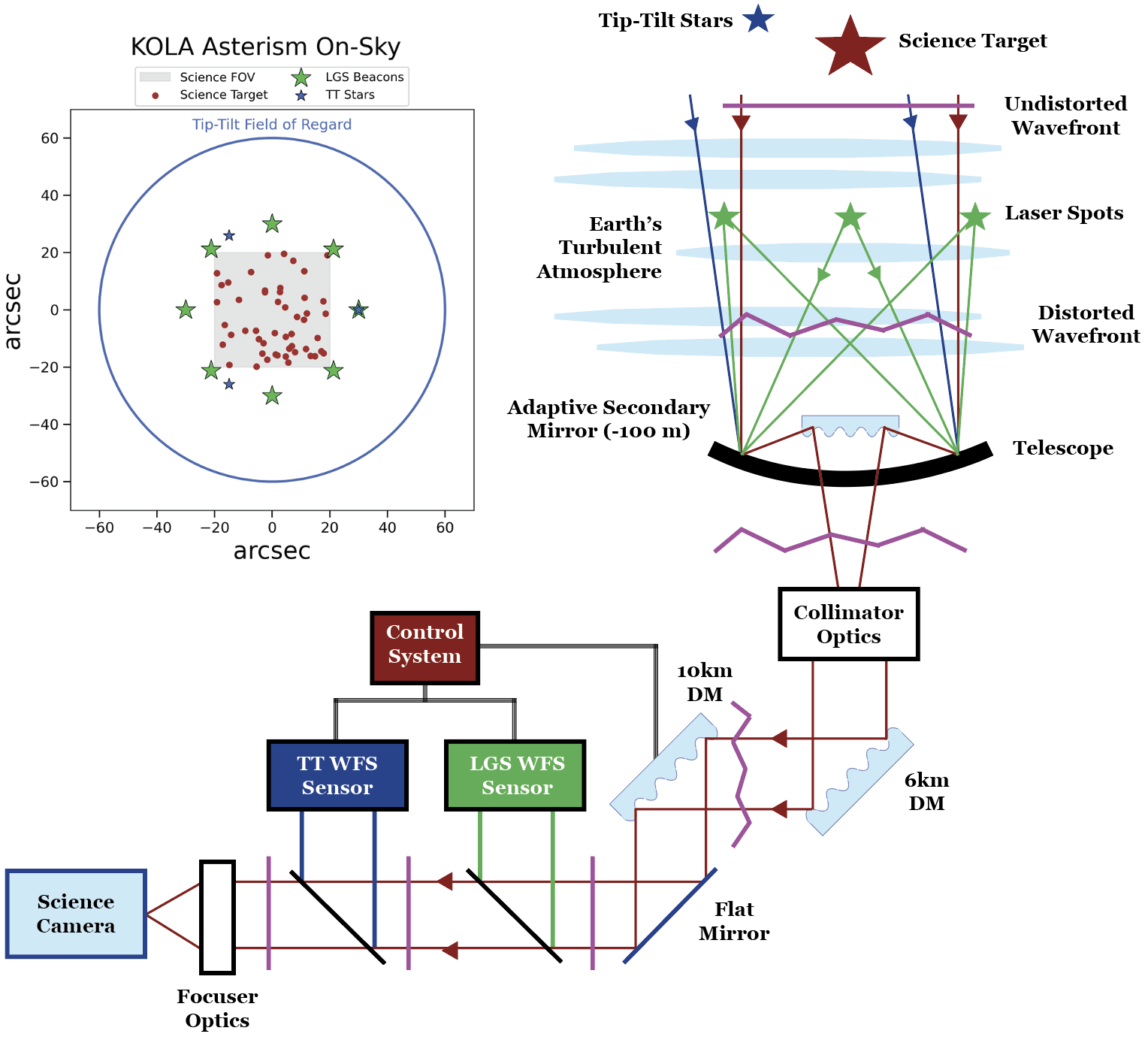}
    \caption{A nominal schematic of the KOLA visible-light multi-conjugate adaptive optics system. An adaptive secondary mirror (conjugated to $-100\,\text{m}$) provides atmospheric wavefront correction, complemented by two high-altitude DMs conjugated to $6\,\text{km}$ and $10\,\text{km}$, respectively. The inset figure shows the nominal on-sky asterism of laser guide stars and tip-tilt stars: eight LGS placed at a $30''$ radius for higher-order correction, and three TTNGS at a $30''$ radius. Performance trade studies in this paper explore varying the number of ASM actuators and the count of laser guide stars.}
    \label{fig:kola_schematic}
\end{figure}

\subsection{Setup and MAOS Configuration}
\label{sec:Setup and MAOS Configuration}

\begin{deluxetable}{c c c c c c c}[t!]
    \tablecaption{MAOS input parameters used to vary the number of $8$th magnitude sodium laser guide stars in the KOLA simulations. Columns 4 and 5 list the angular positions of all three types of WFS (LGS, TT, LB) in arcseconds.}
    \label{tab:lgs_study}
    \tablehead{\colhead{\textbf{\# of LGS}} & \colhead{\textbf{\# of WFS}} & \colhead{\textbf{WFS $x$-position ($''$)}} & \colhead{\textbf{WFS $y$-position ($''$)}} & \colhead{\textbf{LGS $x$ (m)}} & \colhead{\textbf{LGS $y$ (m)}}}
    \tablewidth{\linewidth}
    \startdata
        6 &  [6 3 1] &  [30 15 -15 -30 -15 & [0 25.98 25.98 0 -25.98 & [1.0 0.5 -0.5 -1.0 & [0 0.87 0.87 0  \\
        & & 15 30 -15 -15 0] & -25.98 0 25.98 -25.98 0] &  -0.5 0.5]*6.5 & -0.87 -0.87]*6.5 \\
    \tableline
        7 &  [7 3 1] & [30 18.70 -6.68  & [0 23.45 29.25 13.02 & [1 0.62 -0.22 & [0 0.78 0.97 0.43 \\
        & & -27.03 -27.03 -6.68  & -13.02 -29.25 -23.45 & -0.9 -0.9 -0.22  & -0.43 -0.97  \\ 
        & & 18.70 30 -15 -15 0] & 0 25.98 -25.98 0] & 0.62]*6.5 & -0.78]*6.5 \\ 
    \tableline
        8 & [8 3 1] & [30 21.21 0 -21.21 & [0 21.21 30 21.21 0 & [1 0.71 0 & [0 0.71 1 \\ 
        & & -30 -21.21 0 21.21 & -21.21 -30 -21.21 & -0.71 -1 -0.71 & 0.71 0 -0.71 \\ 
        & & 30 -15 -15 0] & 0 25.98 -25.98 0] & -0 0.71]*6.5 & -1 -0.71]*6.5  \\
    \tableline
        9 & [9 3 1] & [30 22.98 5.21 -15 & [0 19.28 29.54 25.98 & [1 0.77 0.17 & [0 0.64 0.98 \\ 
        & & -28.19 -28.19 -15 & 10.26 -10.26 -25.98 & -0.5 -0.94 & 0.87 0.34 \\ 
        & & 5.21 22.98 30 & -29.54 -19.28 0 & -0.94 -0.5 & -0.34 -0.87 \\
        & & -15 -15 0] & 25.98 -25.98 0] & 0.17 0.77]*6.5 & -0.98 -0.64]*6.5 \\
    \tableline
        10 & [10 3 1]& [30 24.27 9.27 -9.27 & [0 17.63 28.53 28.53 & [1 0.81 0.31 & [0 0.59 0.95 \\ 
        & & -24.27 -30 -24.27 & 17.63 0 -17.63 & -0.31 -0.81 -1 & 0.95 0.59 0.0 \\
        & & -9.27 9.27 24.27 & -28.53 -28.53 -17.63 0 & -0.81 -0.31 & -0.59 -0.95 \\
        & & 30 -15 -15 0] & 25.98 -25.98 0] & 0.31 0.81]*6.5 & -0.95 -0.59]*6.5 \\
    \enddata 
\end{deluxetable}

\begin{deluxetable}{c c c}[t!]
    \tablecaption{MAOS input parameters used to vary the actuator count on the adaptive secondary mirror.}
    \label{tab:act_study}
    \tablehead{
        \colhead{\textbf{Actuator}} & \colhead{\textbf{Actuator}} & \colhead{\textbf{Stellar Flux}} \\
        \colhead{\textbf{Count}} & \colhead{\textbf{Spacing (m)}} & \colhead{\textbf{($e^-/\text{subap}$)}}
    }
    \tablewidth{0pt}
    \startdata
        1000 & 0.308 & 131 \\
    \tableline
        1250 & 0.276 & 105 \\
    \tableline
        1500 & 0.252 & 87 \\
    \tableline
        1750 & 0.233 & 75 \\
    \tableline
        2000 & 0.218 & 65 \\
    \tableline
        2250 & 0.206 & 58 \\
    \tableline
        2500 & 0.195 & 52 \\
    \tableline
        2750 & 0.186 & 47 \\
    \tableline
        3000 & 0.178 & 43 \\
    \tableline
        3250 & 0.171 & 40 \\
    \tableline
        3500 & 0.165 & 37 \\
    \tableline
        3750 & 0.159 & 35 \\
    \tableline
        4000 & 0.154 & 32 \\
    \tableline
        4250 & 0.150 & 31 \\
    \tableline
        4500 & 0.145 & 29 \\
    \tableline
        4750 & 0.141 & 27 \\
    \tableline
        5000 & 0.138 & 26 \\
    \tableline
        5250 & 0.135 & 25 \\
    \tableline
        5500 & 0.131 & 23 \\
    \tableline
        5750 & 0.129 & 23 \\
    \tableline
        6000 & 0.126 & 21 \\
    \enddata 
    \tablecomments{The actuator spacing is given in equivalent separation on the primary mirror.}
\end{deluxetable}

The baseline KOLA configuration begins with the validated Keck pupil ($11.0\,\text{m}$ primary, $2.4\,\text{m}$ secondary obscuration; see Table \ref{tab:vismcao} in Appendix \ref{appendix:MAOS Parameters for KOLA Performance Simulations}) and extends it to a three-mirror MCAO system. 
The effective primary mirror diameter reflects the size of the pupil mask image.  
The DM conjugation heights are set with the parameter \texttt{dm.ht}: the ASM is conjugated to $-100\,\text{m}$, and two additional mirrors are placed at $6\,\text{km}$ and $10\,\text{km}$.
The actuator spacing on the DMs, relative to the Keck primary mirror in meters, is set by \texttt{dm.dx}.
In the performance trade studies, we modify \texttt{dm.dx} for the ASM, but leave the parameter constant for the high-altitude downstream deformable mirrors at $\sim3200$ actuators (reflecting the design of the HAKA DMs \citep{10.1117/12.3020355}. 
Additionally, we increase the actuator stroke and inner-actuator stroke for the DMs (\texttt{dm.stroke} and \texttt{dm.iastroke}) for all three mirrors relative to the current Keck AO system.
For the ASM, it has a stroke of $30\,\mu\text{m}$ with an inner-actuator stroke of $2.0\,\mu\text{m}$.
These parameters help increase the field of view, the ability to correct large wavefront errors and increase the throughput to downstream instruments like spectrographs.

For wavefront sensing, the number of sensors is defined by \texttt{powfs.nwfs}.
The nominal KOLA configuration uses ten LGS wavefront sensors (WFS), three TTNGS WFS, and one low-bandwidth (LB) ``truth" wavefront sensor. 
Relative to the current SCAO configuration, the LGS WFS detectors are upgraded from $\sim3\,\text{e}^-$ read noise per pixel to $0.5\,\text{e}^-$, and sampling per subaperture increases from 4x4 to 5x5 pixels. 
These changes improve slope accuracy and stability at short wavelengths.
The tip-tilt sensors are shifted from R-band to H-band to increase sensitivity to fainter stars, with a target brightness of $H<12\,\text{magnitude}$ within $60''$ of the field center (see KOLA schematic, Figure \ref{fig:kola_schematic}). 

The guide-star geometry consists of multiple sodium beacons arranged on a $30''$ radius ring (side-launch, $90\,\text{km}$ altitude, geometric elongation modeled) and three natural guide stars placed at the same $30''$ radius to sense tip-tilt and low order modes.
As explained in more detail in Section \ref{subsec:Actuator and LGS Trade Study Configuration}, during the LGS count trade study, the number of LGS is varied between 6 and 10, all evenly spaced on a $30''$ radius. 
To focus on the tomographic effects, telescope-induced aberrations and vibration terms (\texttt{surf} and \texttt{wspsd}) are set to zero in the KOLA runs.

All simulation parameters and assumptions used for these runs are summarized in the KOLA MCAO parameters table in the Appendix A), and the system geometry is shown in the KOLA schematic (Figure \ref{fig:kola_schematic}). 
These inputs form the basis of the actuator and LGS trade studies that follow. 

\subsection{Actuator and LGS Trade Study Configuration}
\label{subsec:Actuator and LGS Trade Study Configuration}

To explore the design space for KOLA, we conducted a set of trade studies varying two of the most critical system parameters: the number of actuators on the adaptive secondary mirror and the number of sodium laser guide stars. 
In the early phases of performance simulations, we saw the largest improvement in wavefront error by increasing the number of sodium laser guide stars. 
The magnitude of the sodium LGS had more of a second-order effect on the high-order wavefront error.

For the ASM, actuator counts were varied from 1000 to 6000 in increments of 250, covering the range from coarse to densely packed actuator grids. 
In MAOS, actuator spacing across the primary pupil is set by the parameter \texttt{dm.dx}, with smaller values corresponding to higher actuator densities. 
As the spacing decreases, finer spatial structures in the incoming wavefront can be corrected.
To maintain consistent coupling between the deformable mirror and the wavefront sensor sampling, the subaperture geometry was scaled automatically for each configuration.
The stellar flux per subaperture (\texttt{powfs.siglev}, measured in electrons per subaperture per step) was adjusted for each configuration of actuators while a low constant background (\texttt{powfs.bkgrnd} = $0.1\,\text{e}^-/\text{subap}$) was used throughout.
Table \ref{tab:act_study} summarizes these parameters: the first column lists the total number of ASM actuators, the second column gives the actuator spacing in meters (\texttt{dm.dx}), and the third reports the corresponding stellar flux (\texttt{powfs.siglev}).

For the LGS configuration, we tested asterism containing 6 to 10 sodium beacons, each modeled as an $8$th magnitude source (equivalent to a $22\,\text{W}$ laser projected to $90\,\text{km}$ altitude). 
Because the apparent brightness of an LGS spot depends on the laser power, beam quality, and density and structure of the mesospheric sodium layer, we adopt a baseline conversion such that an $8.1$-magnitude LGS corresponds to a $20\,\text{W}$ laser. 
The scaling between LGS magnitude and laser power is given by
\begin{equation}
    P(m) = P_{\mathrm{ref}} \, 10^{-0.4\,(m - m_{\mathrm{ref}})} ,
\end{equation}

where $P(m)$ is the launch power and $P_{\text{ref}}$, $m_{\text{ref}}$ provide the reference pair.

The total number of wavefront sensors (\texttt{powfs.nwfs}) was updated for each LGS asterism.
All simulations contained three TTNGS WFS fixed on a $30''$ radius, and a single on-axis low-bandwidth sensor for truth sensing.
The angular positions of each WFS are set in MAOS using \texttt{wfs.thetax} and \texttt{wfs.thetay}, while \texttt{powfs0\_llt.ox} and \texttt{powfs0\_llt.oy} define the launch telescope coordinates relative to the Keck primary mirror. 
These parameters were recalculated at each step to maintain a uniform $30''$ circular asterism.
Table \ref{tab:lgs_study} lists these values: the first column gives the total number of sodium LGS in the asterism, the second (\texttt{powfs.nwfs}) shows the total number of wavefront sensors (LGS, TTNGS, and LB), the third and fourth columns (\texttt{wfs.thetax}, \texttt{wfs.thetay}) lists the angular positions of all three types of sensors in arcseconds, and the final two columns (\texttt{powfs0\_llt.ox}, \texttt{powfs0\_llt.oy}) specify the laser launch locations. 

The results of these trade studies are presented in \ref{subsec:Performance Simulation Results}, with the corresponding performance metrics: Strehl ratio, empirical FWHM, and encircled energy radius which guide the selection of KOLA's nominal system configuration.

\subsection{Performance Simulation Results}
\label{subsec:Performance Simulation Results}

\begin{figure}[t]
    \centering
    \includegraphics[width=\linewidth]{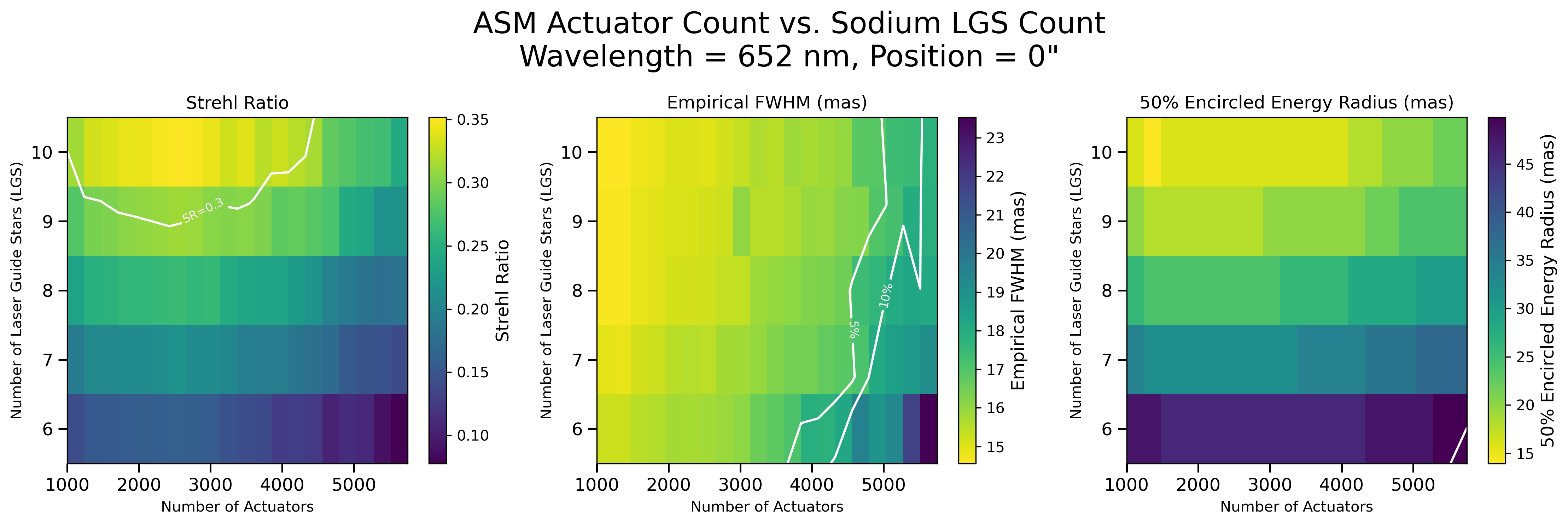}
    \caption{ASM actuator count and number of sodium LGSs trade study for a fixed wavelength ($652\,\text{nm}$) and a fixed position (on-axis). From left to right: Strehl ratio, empirical full-width at half-max (mas) and 50\% of the encircled energy radius (mas). }
    \label{fig:three_metrics}
\end{figure}

\begin{figure}[t]
    \centering
    \includegraphics[width=\linewidth]{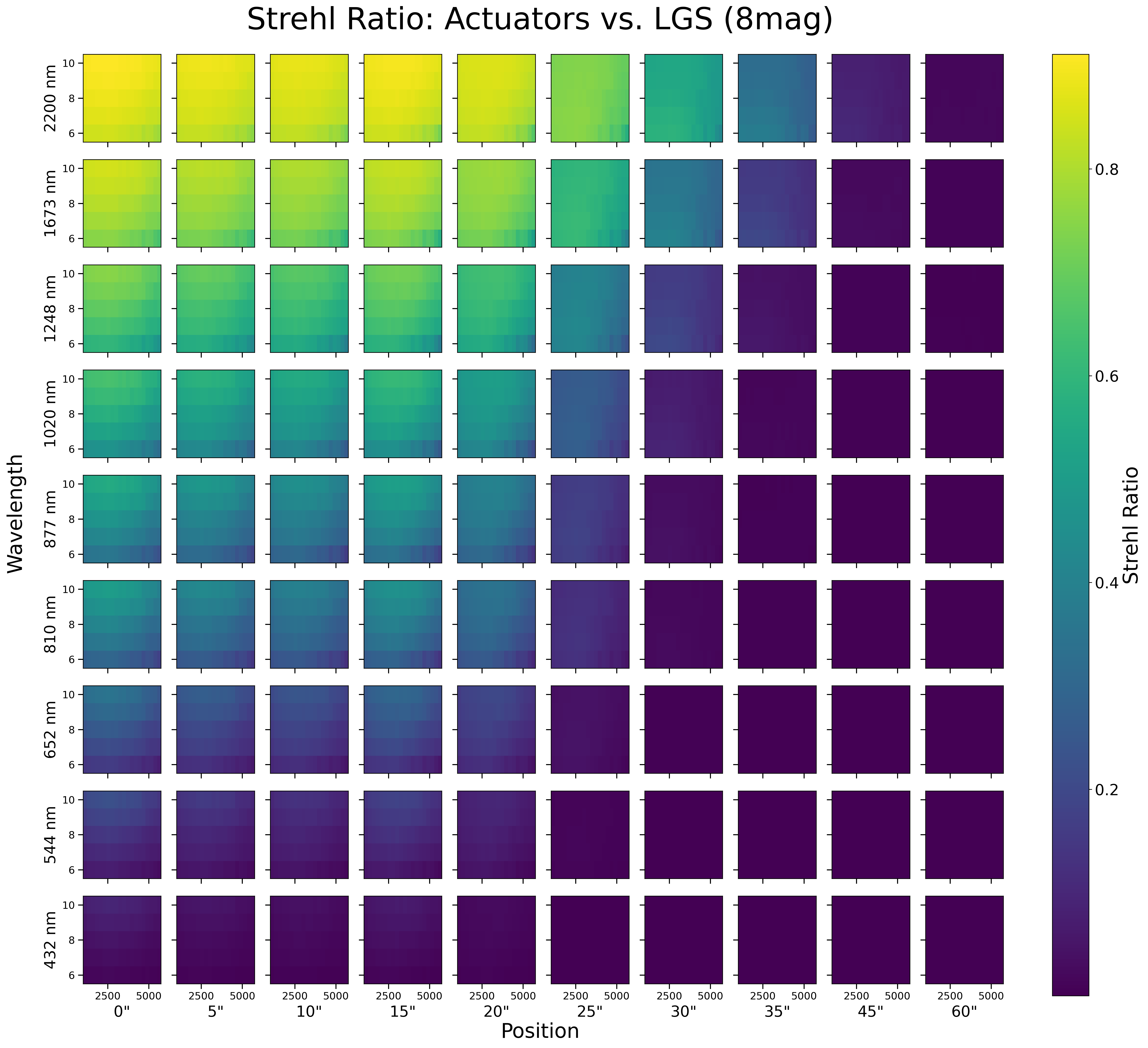}
    \caption{Each panel shows the actuator count on the adaptive secondary mirror versus the number of sodium laser guide stars at a radius of $30''$. The grid spans 9 optical/near-infrared wavelengths ($432\,\text{nm}$ to $2200\,\text{nm}$) and 10 field positions, from on-axis ($0''$) out to $60''$. Colors represent the Strehl ratio, with values near 1 in yellow and values near 0 in purple. All sodium LGS are set to 8th magnitude, while the three TTNGS remain fixed at magnitude 8 with positions at a radius of $30''$ throughout the study.}
    \label{fig:act_vs_lgs_wvl_pos}
\end{figure}

\begin{figure}[t!]
    \centering
    \includegraphics[width=0.5\linewidth]{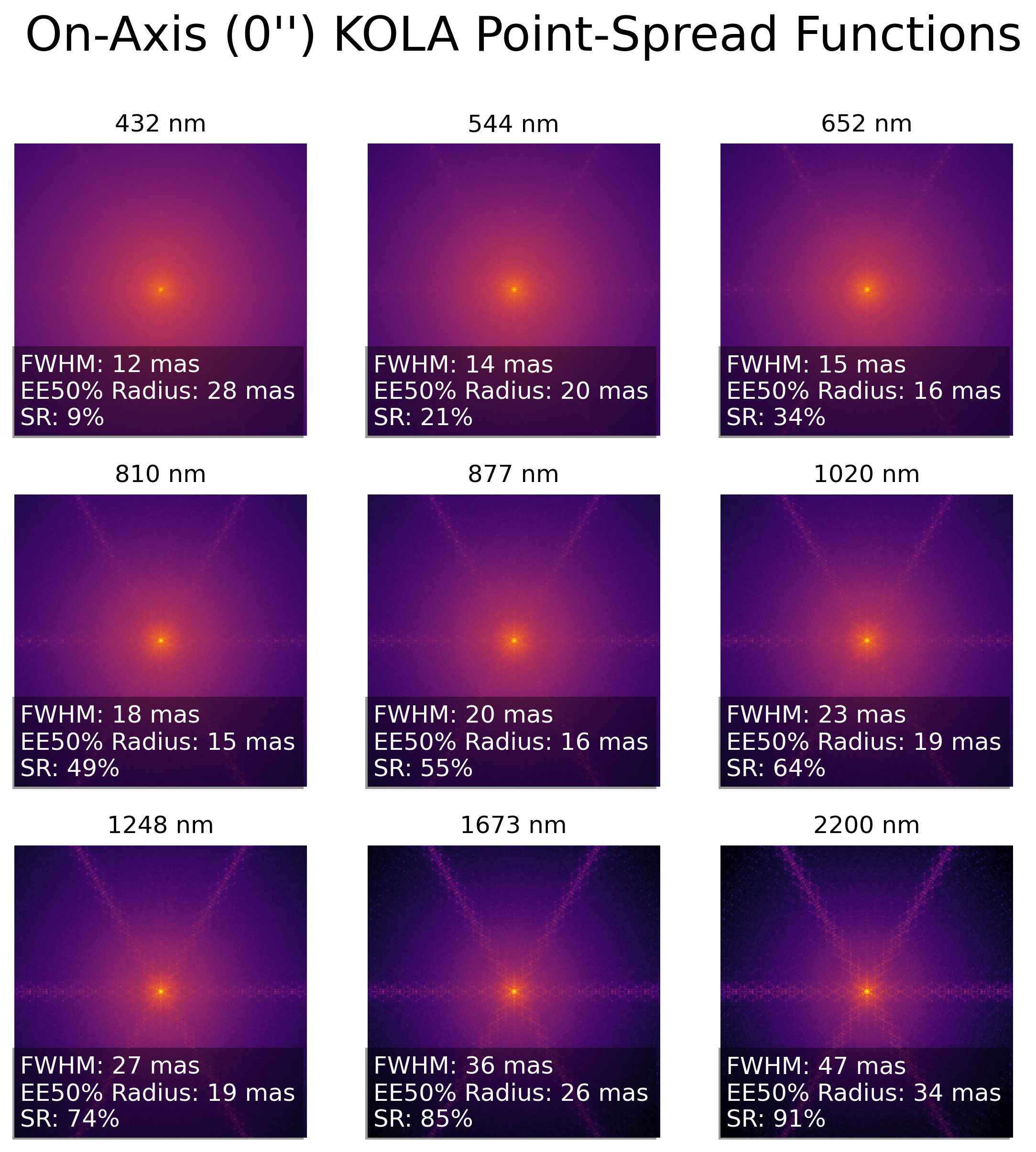}
    \caption{Point-spread functions across several wavelengths for the nominal KOLA design with 3600 actuators and ten sodium laser guide stars at a radius of 30''. All PSFs correspond to the on-axis position ($0''$) and are shown with logarithmic intensity scaling.}
    \label{fig:psfs_at_wvl_metrics}
\end{figure}

\begin{figure}[t!]
    \centering
    \includegraphics[width=\linewidth]{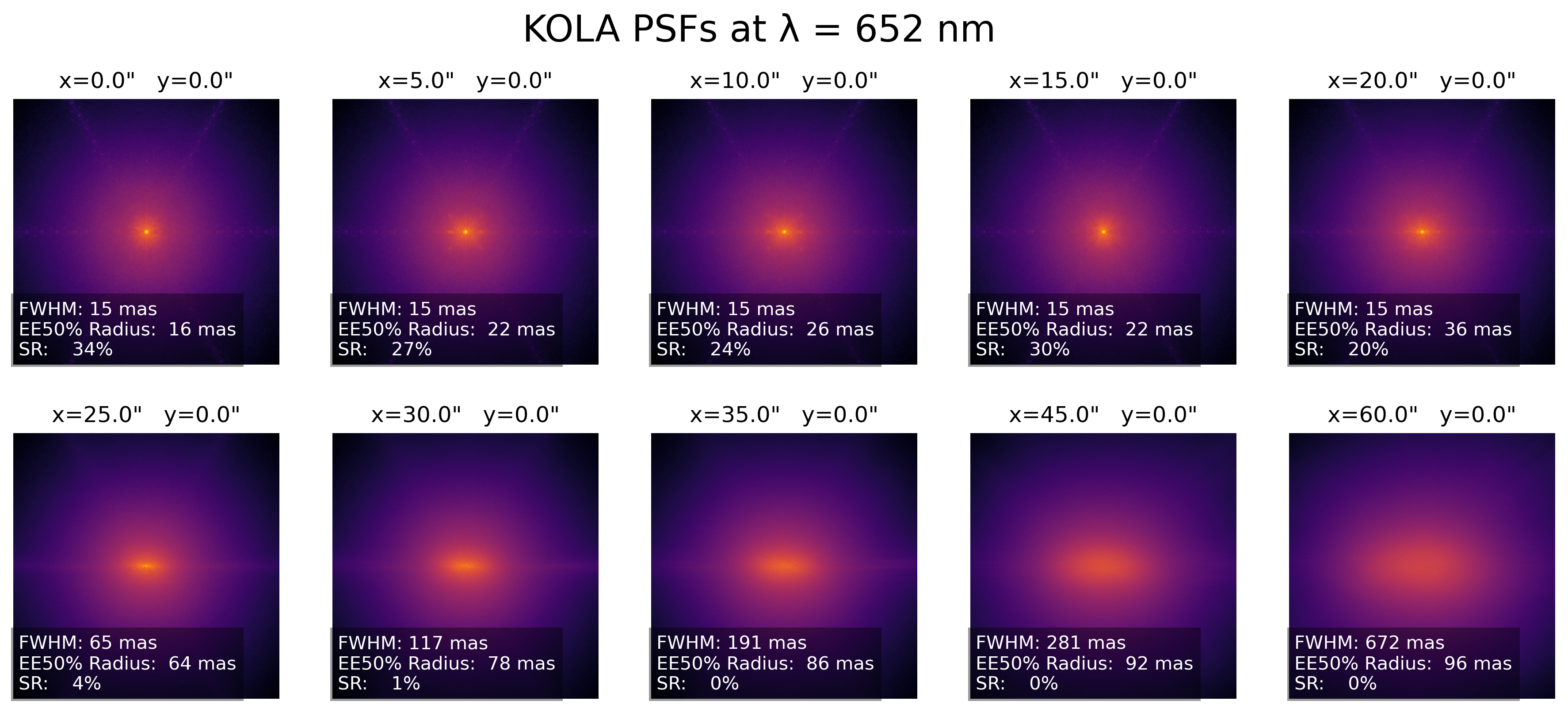}
    \caption{Point-spread functions across a field of view for the nominal KOLA for the same configuration as Figure \ref{fig:psfs_at_wvl_metrics}. All PSFs correspond are at a wavelength of $652\,\text{nm}$ and are shown with logarithmic intensity scaling.}
    \label{fig:psfs_at_652nm_metrics}
\end{figure}

\clearpage

Figure \ref{fig:three_metrics} shows the Strehl ratio, empirical full-width at half maximum and 50\% of the encircled energy radius for several combinations of the performance for an on-axis $652\,\text{nm}$ KOLA system. 
The $x$-axis is the number of actuators on the adaptive secondary mirror ranging from 1000 to 6000 in increments of 250.
The $y$-axis is the number of sodium laser guide stars placed uniformly on a circle with a radius of $30''$.
The white lines correspond to science requirements for KOLA. 
On the Strehl ratio plot (left), there is a white line demarcating where KOLA would achieve a Strehl ratio greater than 30\%.
On the FWHM plot (middle), the white lines depict where KOLA reaches 5\% and 10\% of the diffraction limit at the $652\,\text{nm}$ wavelength.
A higher quality value is represented by yellow for all three of the plots, and each of the plots show a trend toward the top-left corner corresponding to lower actuator counts and higher laser guide star counts. 

Figure \ref{fig:act_vs_lgs_wvl_pos} shows the configurations for KOLA across several positions in the field of view and the wavelength in terms of the Strehl ratio.
The wavelengths range from $432\,\text{nm}$ to $2200\,\text{nm}$ from on-axis $0''$ to $60''$. 
Higher Strehl ratios are occurring (yellow) in the top-left corner of the map corresponding to longer wavelengths and on-axis.

\subsection{Nominal Design for KOLA}
\label{sec:Nominal Design for KOLA}

We present a nominal design for KOLA consisting of ten sodium lasers, each 8th magnitude (22W), placed on a $30''$ radius and 3600 actuators on the adaptive secondary mirror, together will all parameters also defined in the baseline KOLA setup.
This configuration achieves a FWHM of $15\,\text{mas}$ and a Strehl ratio of 34\% at $652\,\text{nm}$ on-axis. 
Figure \ref{fig:psfs_at_wvl_metrics} displays the PSFs and corresponding Strehl ratio, FWHM and 50\% of the encircled energy radius for the nominal KOLA configuration on-axis at different wavelengths. 
Figure \ref{fig:psfs_at_652nm_metrics} shows the same PSF metrics as Figure \ref{fig:psfs_at_wvl_metrics} but at a single wavelength ($652\,\text{nm}$) across a field of view.

Although $3600$ actuators on the ASM is not the absolute optimum according to Figure \ref{fig:psfs_at_wvl_metrics} for 10 sodium LGS, this choice provides margin against variable seeing conditions preserves flexibility for future upgrades.

\section{Discussion and Conclusions}
\label{sec:Discussion and Conclusions}

This nominal design satisfies several of the top-level performance requirements listed in Table~\ref{tab:science_requirements}. 
First, it meets Requirement 1 by delivering AO-corrected PSFs across the full wavelength range from the visible to the near-infrared.
Second, it meets Requirement 2 by achieving spatial resolutions within 10\% of the diffraction limit across all simulated wavelengths.
The configuration also satisfies Requirement 4 by providing uniform correction over a $60''$ diameter field of view, and meets Requirement 5 by delivering Strehl ratios of at least 30\% over a $20''$ field and at least 15\% over a $30''$ field at $650\,\text{nm}$.
Requirements 3 and 6 which address sky coverage at high Galactic latitude and the ability to achieve high Strehl ratios over a 3' field of view for bright TTNGS, will require further work.

The KOLA project is in the conceptual design phase and
we emphasize that the results presented here are the beginning of the design optimization process, not its conclusion. Further simulation work is required to refine and validate the KOLA concept. 
In particular, exploring a broader parameter space (e.g., deformable mirror conjugate heights, actuator counts on high-altitude DMs, factors to match MAOS outputs to the performance of the on-sky Keck AO system, and different LGS asterism diameters in geometries including centered vs. offset configurations) will be essential to converging on an optimal design.

Based on the nominal design, KOLA offers higher Strehl over a larger field of view than MAVIS \citep{2022SPIE12185E..3LA}. However, more comparisons are needed over a wider range of seeing conditions. 
Compared to the GeMS system, KOLA has a smaller field of view by 2x in diameter; but a Strehl ratio that is 2-3$\times$ higher at IR wavelengths. 
More detailed comparisons of the sky coverage, PSF uniformity over the field, and delivered science performance are planned for the future. 

\begin{acknowledgments}
    This simulated adaptive optics performance research for KOLA used resources of the National Energy Research Scientific Computing Center (NERSC), a US Department of Energy Office of Science User Facility located at Lawrence Berkeley National Laboratory, operated under contract No. DE-AC02- 05CH11231 using NERSC awards HEP-ERCAP0026816 and HEP-ERCAP0026816.
    B.P., J.R.L., B.D., and C.A.C.~acknowledge support from the National Science Foundation under grant No.~2108185 and 2511697, the California Association for Research in Astronomy under grant No.~16278 and the Heising-Simons Foundation under grant No.~2022-3542.
    
    Some of the data presented herein were obtained at Keck Observatory, which is a private 501(c)3 non-profit organization operated as a scientific partnership among the California Institute of Technology, the University of California, and the National Aeronautics and Space Administration. The Observatory was made possible by the generous financial support of the W.~M.~Keck Foundation. The authors wish to recognize and acknowledge the very significant cultural role and reverence that the summit of Maunakea has always had within the Native Hawaiian community. We are most fortunate to have the opportunity to conduct observations from this mountain.
\end{acknowledgments}

\software{Numpy \citep{harris2020array}, Matplotlib \citep{Hunter:2007}, Astropy \citep{astropy:2022}, SciPy \citep{2020SciPy-NMeth}, MAOS \citep{wang2018maos}}

\facility{Keck:I}

\appendix

Section \ref{appendix:MAOS Parameters for KOLA Performance Simulations} contains Table \ref{tab:vismcao} which describes the parameters in MAOS updated from the current Keck AO system to KOLA. 
While Section \ref{appenidx:Validation of Current Keck AO System in MAOS} describes the results of the validation simulations with the simulated current Keck AO system compared to on-sky data.

\section{MAOS Parameters for KOLA Performance Simulations}
\label{appendix:MAOS Parameters for KOLA Performance Simulations}

Table \ref{tab:vismcao} contains the parameters for the nominal KOLA configuration. The values for \texttt{dm.stroke} and \texttt{dm.iastroke} are from \cite{Dekany2019}.

\startlongtable
\begin{deluxetable}{c l c c c}
\tablecaption{Differences in parameters in MAOS for the current Keck AO system and the nominal KOLA design. For the \textbf{dm.} parameters, the configuration is adaptive secondary mirror, followed by the first and then second downstream deformable mirror. For the \textbf{powfs.} parameters, the configuration is sodium laser guide star wavefront sensor, tip-tilt natural guide star WFS and the low-bandwidth WFS.}
\label{tab:vismcao}
\tablehead{
\colhead{\textbf{Parameter}} &
\colhead{\textbf{Description}} &
\colhead{\textbf{Location}} & 
\colhead{\textbf{Current Keck AO}} & 
\colhead{\textbf{KOLA}} 
}
\tablewidth{\linewidth}
\startdata
    dm.ht   & Height of atmospheric & [ASM DM1 DM2] & [0] & [-100 6000 10000] \\
            & layer DM is conjugated & & & \\
            & to (m) & & & \\
    \tableline
    dm.dx   & Actuator spacing on & [ASM DM1 DM2] & [0.563] & [0.178 .168 .168] \\ 
            & ASM/DM relative to the & & & \\
            & primary mirror (m) & & & \\
    \tableline
    dm.stroke & Surface stroke for & [ASM DM1 DM2] & [2.0] & [30.0 7.0 7.0] \\ 
            & actuators ($\mu$m) & & & \\
    \tableline
    dm.iastroke & Inner-actuator surface & [ASM DM1 DM2] & [1.20] & [2.0 2.0 2.0] \\ 
            & stroke ($\mu$m) & & & \\
    \tableline
    powfs.rne & WFS read noise per & [LGS TT LB] & [3.0 0.1 1.0] & [0.5 0.1 1.0] \\ 
            & pixel (e$^-$ per pixel) & & & \\
    \tableline
    powfs.dsa & Diameter of WFS sub & [LGS TT LB] & [0.563 10 0.563] & [0.563 5 0.563] \\ 
            & -aperture in 1D (m) & & & \\
    \tableline
    powfs.dtrat & Sampling rate of WFS & [LGS TT LB] & [1 1 1000] & [1 1 100000] \\ 
            & in units of sim.dt & & & \\
   \tableline
   powfs.siglev & Stellar flux per sub- & [LGS TT LB] & [1270 64529 147933] & [44 42388 7] \\ 
        & aperture (e$^-$ per & & & \\ 
        & subap in time sim.dtref) & & & \\
   \tableline
   powfs.bkgrnd & Background per sub- & [LGS TT LB] & [1 0 0.004] &  [0.1 1.1 0.001] \\ 
        & aperture (e$^-$ per subap) & & & \\
    \tableline
    powfs.nearecon & Noise equivalent angle & [LGS TT LB] & [43 1 388.4] & [242 0.3 1200] \\ 
    &  (mas$^2$) & & & \\
    \tableline
    powfs.nwfs & Number of LGS, TTNGS  & [LGS TT LB] & [1 1 1] & [10 3 1] \\
        & and LB wavefront sensors & & & \\
    \tableline
   powfs.dx & Sampling of computational & [LGS TT LB] & [1/64 1/16 1/32] & [1/128 1/16 1/32] \\ 
        & optical path for each sub- & & & \\
        & aperture, matches atm.dx (m) & & & \\
    \tableline
    powfs.pixtheta & Size of CCD pixels  & [LGS TT LB] & [3 1.5 1.5] & [1 0.06 1.5] \\ 
        & (arcsec) & & & \\
    \tableline
    powfs.pixpsa & Number of pixels per sub- & [LGS TT LB] & [2 2 8] & [5 4 8] \\ 
        & aperture on a side (pixel  & & & \\
        & per subap) & & & \\
    \tableline
    powfs.misregx/y & Misregistration of the & [LGS TT LB] & [0.0021 0 0.0059] & [0.0021 0 0.0059] \\
        & WFS sub-apertures (fraction & & & \\
        & of a sub-aperture) & & & \\
    \tableline
    powfs.wvl & Central wavelength of  & [LGS TT LB] & [0.589e-6  & [0.589e-6  \\
        & WFS (m) & & 0.72e-6 & 1.63e-6 \\
        & & & 0.72e-6] & 0.72e-6] \\
    \tableline
    powfs.gtype\_recon & Type of reconstructor & [LGS TT LB] & [0 0 0] & [0 1 0] \\
        & (0= averaging gradient,  & & & \\
        & 1=Zernike tilt) & & & \\
    \tableline
    wfs.thetax & Location of LGS, TT  & [LGS TT LB] & [0 0 0] & [30 24.27 9.27  \\ 
        & and LB wave-front sensor & & & -9.27 -24.27 -30 \\
        & in x (arcsec) & & & -24.27 -9.27 9.27 \\
        & & & & -15 -15 30] \\
    \tableline
    wfs.thetay & Location of LGS, TT  & [LGS TT LB] & [0 0 0] & [0 17.63 28.53 \\ 
        & and LB wave-front sensor & & & 28.53 17.63 0 \\
        & in y (arcsec) & & & -17.63 -28.53 -28.53 \\
        & & & & -17.63 0 25.98 \\
        & & & & -25.98 0] \\
    \tableline
    powfs0\_llt.ox & Laser launch telescope & [LGS1 LGS2 LGS3 & [0] & [1.0 0.81 0.31  \\ 
        & x-axis distance from  & LGS4 LGS5 LGS6 & & -0.31 -0.81 -1.0 \\
        & optical axis (m) & LGS7 LGS8 LGS9 & & -0.81 -0.31 0.31 \\
        & & LGS10] & & 0.81]*6.5 \\
    \tableline
    powfs0\_llt.oy & Laser launch telescope & [LGS1 LGS2 LGS3 & [0] & [0.0 0.59 0.95 \\ 
        & y-axis distance from & LGS4 LGS5 LGS6 & & 0.95 0.59 0.0 \\
        & optical axis (m) & LGS7 LGS8 LGS9 & & -0.59 -0.95 -0.95 \\
        & & LGS10] & & -0.59]*6.5  \\
    \tableline
    sim.end & Number of time steps & SIM & 800 & 6400 \\ 
    \tableline
    sim.dt & Sampling frequency of & SIM & 1/472 & 1/1500 \\ 
        & simulation (Hz) & & & \\
    \tableline
    sim.dtref & Reference sampling time & SIM & 1/472 & 1/1500 \\ 
        & to set signal level (s)  & & & \\
    atm.dx & Sampling of the atmosphere, & SIM & 1/64 & 1/128 \\
        & needs to match powfs.dx (m) & & & \\
    \tableline
    surf & Inputs a static non- & SIM & ["Keck\_ncpa\_rms & ['' "] \\
        & common aberration path map & & wfe130nm.fits"] & \\
    \tableline
    wspsd & Wind speed and &  SIM & ["PSD\_Keck\_ws20 & [] \\
        & vibration spectrum & & mas\_vib26mas & \\
        & & & \_rad2.fits"] & \\
    \tableline
    evl.wvl & Wavelengths to & SIM & [0.8 1.0 1.25 & [0.432 0.544 0.652  \\
    & evaluate PSF (m) & & 1.65 2.12]*1e-6 & 0.810 0.877 1.020 \\
    & & & 1.248 1.673 2.200]*1e-6 \\
    \tableline
    evl.psfsize & Size of output PSF  & SIM & [256 256 256  & [256 256 256 \\
        & (pix) for numbers of & & 256 256] & 256 256 256 \\
        & wavelengths in evl.wvl & & & 256 256 256] \\
    \tableline
    fit.fov & Reconstructor field to & SIM & 0 & 30 \\
        & average over (arcsec) & & & \\
    \tableline
    fit.thetax & Reconstructor evaluation & FIT & [0] & [0 0.5 0  \\
        & X positions relative to & & & -0.5 0 0.5 \\
        & field of view & & & -0.5 -0.5 0.5] \\
    \tableline
    fit.thetay & Reconstructor evaluation & FIT & [0] & [0 0 0.5 0 \\
        & Y positions relative & & & -0.5 0.5 0.5 \\
        & to field of view & & &  -0.5 -0.5] \\
    \tableline
    fit.wt & Reconstructor weight & FIT & [1] & [4/9 1/9 1/9 \\
        & for each evaluation point & & & 1/9 1/9 1/36 \\
        & & & & 1/36 1/36 1/36] \\
    \tableline
    tomo.svdthres & Threshold of SVD inversion & TOMO & 1e-9 & 1e-8 \\
\enddata
\end{deluxetable}

\section{Validation of Current Keck AO System in MAOS}
\label{appenidx:Validation of Current Keck AO System in MAOS}

Before extending the MAOS framework to KOLA, we first validated MAOS by reproducing the current Keck SCAO system.
To validate, we varied the magnitude of the tip-tilt natural guide star from $8$th magnitude to $18$th magnitude.
The parameters for this validation are shown in Table \ref{tab:magnitude_scao_params} where the first column is the magnitude of the single TTNGS while the second column contains the stellar flux (\texttt{powfs.siglev}) in electrons per subaperture per step.

Table \ref{tab:magnitude_scao} shows the validation comparison between the current Keck AO system simulated in MAOS and on-sky data.
The first column is the R magnitude of the TTNGS, the second and third columns are the results from the MAOS simulations and on-sky data in the K band \citep{vanDam2007_NGWFC}. 

\begin{deluxetable}{c c}
  \tablecaption{Keck SCAO MAOS parameters for WFS flux from different TTNGS magnitudes.}
  \label{tab:magnitude_scao_params}
  \tablehead{
  \colhead{\textbf{TTNGS Magnitude}} & 
  \colhead{\textbf{Stellar Flux} ($e^{-}/\text{subap}$)}}
  \tablewidth{\linewidth}
  \startdata 
    8  & 101787 \\
    9  & 40522 \\   
    10 & 16132 \\
    11 & 6422 \\
    12 & 2556 \\
    13 & 1017 \\
    14 & 405 \\
    15 & 161 \\
    16 & 64 \\
    17 & 25 \\
    18 & 10 \\
  \enddata 
\end{deluxetable}

\begin{deluxetable}{c c c}
  \tablecaption{Keck SCAO MAOS parameters for WFS flux from different TT and LBWFS NGS magnitudes.}
  \label{tab:magnitude_scao}
  \tablehead{
  \colhead{\textbf{R Magnitude}} & \colhead{\textbf{MAOS: K-Band}} & \colhead{\textbf{On-Sky: K-band}} } 
  \tablewidth{\linewidth}
  \startdata 
    10      & 0.39      & 0.35 \\
    11      & 0.37      & N/A \\
    12      & 0.33      & N/A  \\
    13      & 0.29      & N/A \\
    14      & 0.26      & N/A \\ 
    15      & 0.24      & N/A \\
    16      & 0.16      & 0.29 \\ 
    17      & 0.11      & 0.24 \\
    18      & 0.08      & 0.08 \\
  \enddata 
\end{deluxetable}

\clearpage

\bibliography{bib}{}
\bibliographystyle{aasjournalv7}



\end{document}